## Lattice softening and bandgap evolution in transition-metal-doped ferroelectric AlN: a systematic DFT study

Elizaveta Kalika, Ilya Margolin and Anastasia Chouprik*

*Moscow Institute of Physics and Technology (National Research University), Dolgoprudny, 141700 Russia*

*e-mail: chouprik.aa@mipt.ru*

### Abstract

Ferroelectric wurtzite AlN is a CMOS-compatible alternative to oxide ferroelectrics for non-volatile memory, but its coercive field of 8-10 MV/cm approaches dielectric breakdown. Alloying with Sc lowers the axial-to-lateral lattice ratio *c/a* and hence the switching barrier, yet Sc is scarce and costly. We report a systematic spin-polarized density functional theory screening of 3*d*, 4*d* and 5*d* transition metals substituting Al in AlN at 6.25%, 12.5% and 25%. Remanent polarization stays high throughout (1.11-1.43 $C/m^2$), so dopant selection is governed by the *c/a* ratio and the bandgap. Sc gives the largest reduction of *c/a* (-2.85% at the lowest substitution), followed by Nb and Ta, which keep the bandgap open despite heterovalent substitution. The *c/a* ratio of Nb, Ta, Zr and Y evolves non-monotonically, which the density of states traces to localized in-plane *d* states delocalizing into the conduction band. Spin polarization removes the apparent lattice softening of mid-row 3d elements.

## INTRODUCTION

Ferroelectric wurtzite aluminum nitride (w-AlN) has recently gathered significant attention as a highly promising candidate for non-volatile ferroelectric random-access memory (FRAM) applications [1]. Unlike conventional perovskite ferroelectrics, AlN-based films offer high remanent polarization, excellent thermal stability, and compatibility with silicon-based complementary metal-oxide-semiconductor (CMOS) technology, making them ideal for ultra-dense and high-speed memory integration [2]. Furthermore, as a pure nitride semiconductor, the AlN lattice is entirely free of oxygen atoms. This crystal structure prevents the formation and migration of oxygen vacancies, which are the main cause of polarization decay and fatigue in oxide-based ferroelectrics (such as doped $HfO_2$ or $Pb(Zr_xTi_{1-x})O_3$) [3, 4]. Consequently, AlN-based films naturally provide potentially unlimited data retention and long-term device reliability.

However, the main problem preventing the widespread commercialization of pure AlN-based devices is its excessively high intrinsic coercive field $E_c$ ~ 8-10 MV/cm, which approaches the dielectric breakdown voltage of the material [5, 6]. This high switching threshold originates from the geometry of the switching pathway itself [7]. Polarization reversal proceeds not by dipole rotation but by flattening of the $AlN_4$ tetrahedra, the cation passing through a non-polar layered configuration of the h-BN type (space group $P6_3/mmc$) on its way to the mirror-image position; the energy of this configuration relative to the wurtzite ground state sets the switching barrier. The $c/a$ ratio measures how far the equilibrium structure already lies along this pathway: the ideal wurtzite value is 1.633 and that of the fully flattened layered phase is close to 1.2, so a reduced $c/a$ corresponds to tetrahedra that are already partially flattened and to a correspondingly lower barrier. This is the mechanism by which Sc operates (the larger $Sc^{3+}$ ion favours a flatter coordination and expands the lateral parameter a) and it accounts for the near-linear decrease of both $c/a$ and $E_c$ with Sc content in $Al_{1-x}Sc_xN$ [8, 9, 10]. Alternative descriptions of the switching pathway have been proposed, based on the cation coordination environment and the ionicity of the bonding [11] or on an individual switching mode through a β-BeO-like intermediate [12]. The descriptor used here does not depend on which pathway dominates, since all of them pass through a flattened, non-polar configuration whose energy falls as $c/a$ decreases. It should be noted that the barrier obtained from such structural arguments corresponds to homogeneous switching of the whole cell, whereas experimental coercive fields are governed by nucleation and domain wall motion and are typically lower; the $c/a$ ratio is therefore used here as a descriptor of the relative tendency of a dopant to soften the lattice, rather than as a quantitative predictor of $E_c$.

To date, alloying AlN with scandium has been the standard approach to lower the $c/a$ ratio and mitigate the coercive field. Nevertheless, scandium is a rare, chemically reactive, and extremely expensive transition metal, which raises serious concerns regarding the cost-effectiveness and supply chain sustainability of Sc-doped AlN devices in industrial manufacturing [13]. As an alternative, incorporating lighter group-III elements, such as boron to form $Al_{1-x}B_xN$ thin films, has recently been explored [14]. While $Al_{1-x}B_xN$ preserves a wide energy bandgap, it frequently exhibits incomplete wake-up [15] and increased current leakage during electrical cycling [16]. Yttrium has likewise been proposed as a cheaper isovalent substitute [17]. Existing computational work, however, has either compared a few candidates such as Sc, Y and B [12, 17], screened a broad set of dopants against a single target property [11], or surveyed dopants added to an already Sc-containing host [18]; the full transition metal series in undoped AlN has not been assessed against lattice softening, polarization and insulating character taken together.

In this work, we employ comprehensive density functional theory (DFT) screening to systematically investigate the effect of substituting aluminum with a wide range of 3d, 4d, and 5d transition metals. We analyze the evolution of the $c/a$ ratio, remanent polarization, and electronic bandgap across three distinct doping concentrations: 6.25%, 12.5%, and 25%. Our results reveal highly non-linear and unexpected oscillatory structural behaviors at higher doping regimes, offering critical insights into the selection of optimal cost-efficient candidates for next-generation FRAM technologies.

## COMPUTATIONAL DETAILS

All density-functional calculations were performed using Quantum ESPRESSO package [19]. The exchange-correlation interactions were treated using the Generalized Gradient Approximation (GGA) in the Perdew-Burke-Ernzerhof (PBE) formulation [20]. The ion-electron interactions were described via Projector Augmented-Wave (PAW) pseudopotentials [21].

A 2 × 2 × 2 supercell of 32 atoms was used to construct pure AlN and substituted AlN wurtzite structures. To find the optimal atomic configurations for multi-atom substitutions (at 12.5% and 25% concentrations), a systematic screening of symmetry-independent structures was performed. We used the Pymatgen package [22] to generate all theoretically possible spatial arrangements of the dopant atoms within the 2 × 2 × 2 AlN supercell. Using the built-in symmetry analysis tools, the total number of structures was filtered down to a complete set of symmetrically non-equivalent configurations. Structural relaxation calculations were then performed for each of these unique configurations. Finally, the structure possessing the absolute minimum total energy was selected as the most stable ground-state configuration for all further calculations.

Variable-cell structural relaxations were carried out until the total forces on atoms were smaller than 1.5 meV/A and the total energy converged within $7\times10^{-5}$ eV. For these structural optimization steps, a 6 × 6 × 6 Monkhorst-Pack *k*-point grid and a kinetic-energy cutoff of 1090 eV was used. In polarization and DOS calculations, we set a denser grid of 10 × 10 × 10 *k*-points mesh for the Brillouin zone sampling. V, Ni and W could not be converged reliably at any of the substitution levels considered and are excluded from the analysis. In addition, individual configurations of Cr (25%) and of Rh and Re (12.5%) did not converge within the criteria adopted here; these are omitted and appear as blank entries in the Supplementary Material.

For systems containing magnetic 3*d* transition metals with partially filled d-shells (Cr, Mn, Fe, and Co), all calculations were performed using the spin-polarized formalism [23]. To ensure convergence to a ferromagnetic solution, an initial total magnetization was imposed on the supercells, scaled proportionally to the concentration of the magnetic ions (from 3-5 $\mu_B$ at 6.25% up to 12-20 $\mu_B$ at 25% substitution); antiferromagnetic orderings were not considered in the present screening. For the remaining dopants spin-polarized test calculations converged to vanishing magnetic moments and the non-spin-polarized results are reported. Hubbard U corrections were not applied. The DFT+U scheme addresses the residual self-interaction of localized *d* electrons, whereas the effect examined here is the exchange splitting of partially occupied *d* shells, which is captured by the spin-polarized formalism itself. Since the present study compares relative trends across the transition metal series rather than absolute energies, applying a single U value to chemically dissimilar dopants would introduce an arbitrary element-dependent bias; U has instead been used elsewhere as a deliberate tuning parameter for the ionic character of the bonding [11], which is not the purpose of the present screening.

## STRUCTURAL AND FERROELECTRIC MAPPING: C/A RATIO AND POLARIZATION

The main structural descriptor for lowering the coercive field in wurtzite-based ferroelectrics is the axial-to-lateral lattice parameter ratio *c/a*. Pure w-AlN has a high *c/a* ratio of 1.61, leading to a deep potential energy barrier during polarization switching. Substituting Al with transition metals modifies this local structure and can flatten the potential energy landscape [24].

To evaluate the effectiveness of different dopants, a density functional theory-based study was performed across the 3*d*, 4*d* and 5*d* elements. The evolution of the lattice geometry and ferroelectric properties is demonstrated in Figure 1, where the cell color corresponds to the polarization value of the 6.25%-doped AlN and the bars represent the relative change of the *c/a* ratio compared to pure AlN.

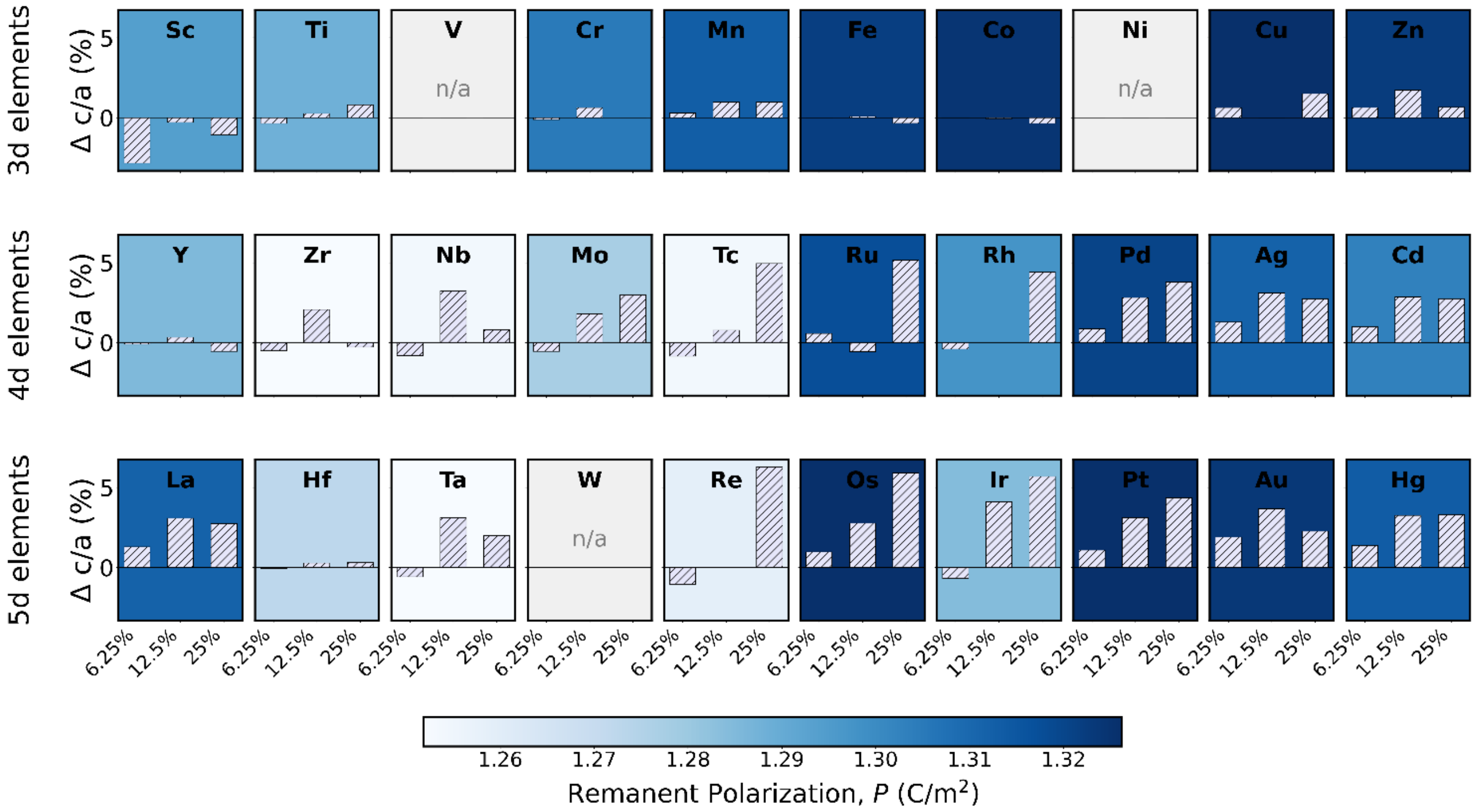

**Figure 1.** Effect of transition metal dopants in wurtzite AlN across 6.25%, 12.5%, and 25% substitution levels. The columns follow the groups of the periodic table, and the rows represent the 3*d*, 4*d* and 5*d* series. Inside each cell, the bars indicate the percentage change in the *c*/*a* ratio (Δ*c*/*a*, %) relative to pristine AlN, where downward bars represent favorable lattice compression. The background color of each cell corresponds to the magnitude of the remanent polarization with 6.25% doping. Missing bars or empty cells denote calculations that did not converge due to structural instability.

It is crucial to note that across all stable configurations and concentrations, the remanent polarization of these AlN-based structures remains remarkably high, bounded within the 1.2-1.3 C/m$^2$ range. Such polarization is an order of magnitude higher than the theoretical values predicted for conventional ferroelectrics, which typically range from 0.5 to 0.7 C/m$^2$ for hafnium oxide [25,26,27] and from 0.5 to 0.8 C/m$^2$ for classic perovskite materials [28,29,30]. As remanent polarization remains consistently massive, its slight variations are less critical for device functionality. Therefore, to ensure visual clarity and avoid an over-cluttered color scheme, the background color gradient in Figure 1 is mapped strictly to the polarization values at the initial 6.25% concentration. Consequently, the engineering priority for low-voltage FRAM applications shifts entirely toward optimizing and minimizing the *c/a* lattice ratio rather than monitoring the polarization magnitude.

As shown in Figure 1, the structural and ferroelectric responses are highly dependent on the specific chemical group and doping concentration. At a low doping concentration of 6.25%, several different trends emerge. Scandium serves as the benchmark dopant, providing a strong reduction in the *c/a* ratio of -2.85% while maintaining remanent polarization $P_r$ ~1.29 C/m$^2$. This structural softening happens because Sc tends to favor a metastable layered hexagonal phase without destroying the ferroelectric double-well potential [24].

Beyond scandium, early transition metals such as Ti, Zr, Nb, Hf and Ta also successfully reduce the *c/a* ratio at 6.25% substitution. Notably, Nb and Ta achieve significant lattice compression (-0.83% and -0.58%, respectively) while holding the polarization at a robust level ($P_r$ ~ 1.25 C/m$^2$).

In contrast, moving rightward across the periodic table toward late transition metals (like Cu, Zn, Ag and Au) leads to a growth of the *c/a* ratio up to +1.90% for Au. Furthermore, this geometric expansion is accompanied by a drop in polarization value. This combination makes late transition metals entirely unsuitable for application in FRAM.

As the substitution concentration increases to 12.5% and 25%, the structural and ferroelectric response of the AlN matrix becomes highly non-linear. The bars in Figure 1 clearly highlight an oscillatory evolution of the *c/a* ratio for early 4*d* and 5*d* dopants. While Nb and Ta compress the lattice at 6.25%, a higher concentration of 12.5% causes a massive lattice distortion, indicating a huge *c/a* expansion of +3.21% for Nb and +3.11% for Ta. However, when the substitution level reaches 25%, this trend completely inverts. The *c/a* ratio drops to 1.616 for Nb and 1.635 for Ta. Notably, this calculated structural trend has been previously experimentally observed via XRD for Ta-substituted AlN thin films [27]. A similar wave-like trend is observed for Zr and Y, which peak at 12.5% and then demonstrate compression at 25% substitution.

In contrast, certain elements affect the *c*/*a* ratio in a unidirectional manner, either increasing or decreasing it. For example, Pd, Ag, Cd, La, Os, Pt, Au and Hg raise the *c*/*a* ratio, which consequently increases the coercive field. This effect is generally undesirable for FRAM devices.

Spin polarization qualitatively changes the structural response of the mid-row 3*d* dopants. When the local moments on Cr, Fe and Co are allowed to develop, the c/a ratio remains essentially that of pristine AlN, and for Mn it increases; the pronounced lattice softening obtained for these elements in non-spin-polarized calculations is therefore spurious. A reduction of c/a survives only for Fe and Co at 25% substitution, where the bandgap has already closed and the material is no longer an insulator. Mid-row 3*d* elements thus offer no usable window for coercive-field engineering in AlN, and screening studies that treat these dopants without spin polarization are liable to overestimate their promise.

## BANDGAP ANALYSIS

To systematically evaluate the insulating performance across all 30 screened structures, the evolution of the effective electronic bandgap ($E_g$) as a function of doping concentration is shown in Figure 2. Further study of the band gap values is necessary, as band gap narrowing may cause large leakage currents in FRAM devices, making them non-functional even if the *c/a* ratio in the ferroelectric layer is reduced.

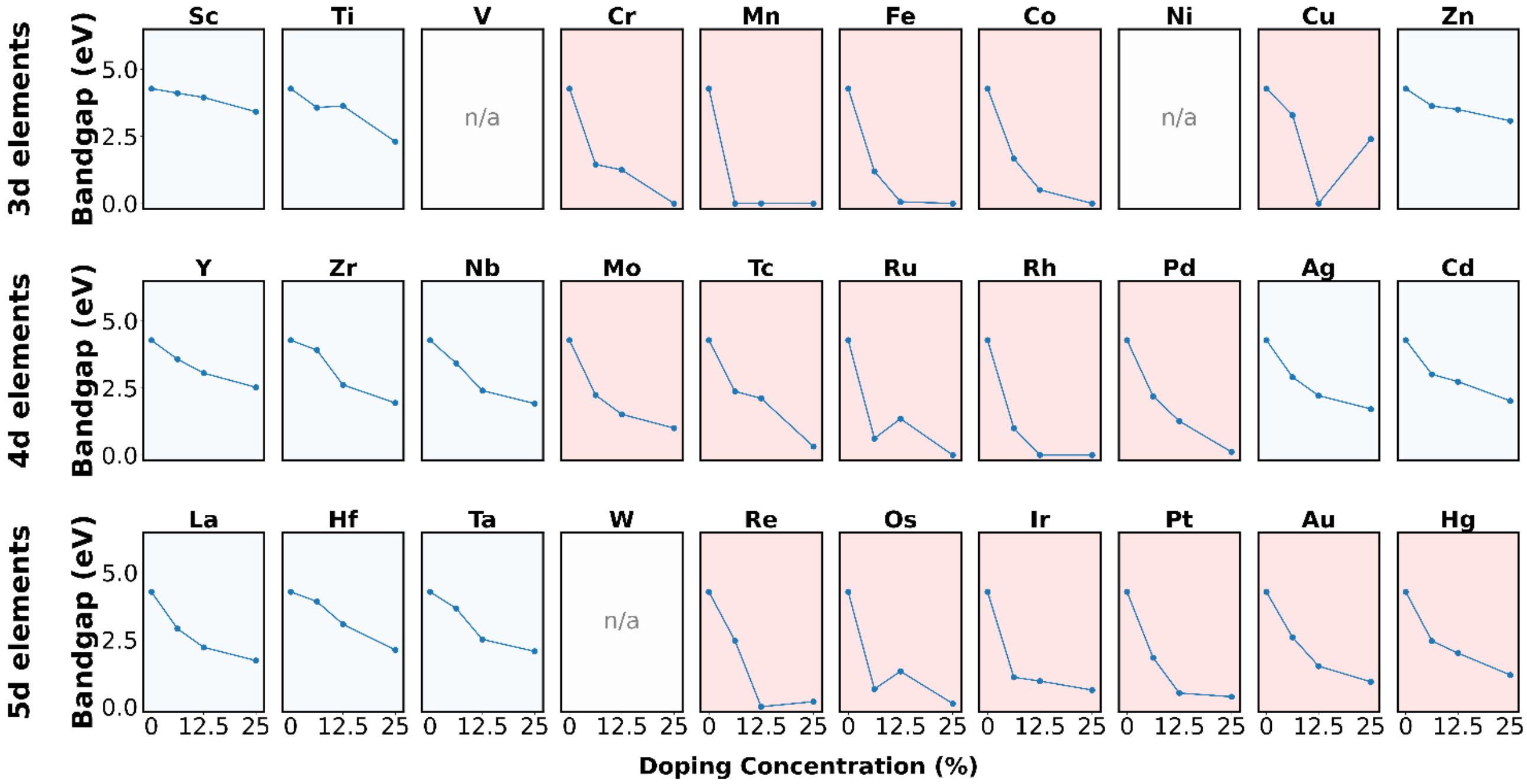

**Figure 2.** Electronic bandgap ($E_g$) as a function of doping concentration for the studied 3*d*, 4*d* and 5*d* transition metals in w-AlN. Cells highlighted in pink indicate narrow-gap semiconductor or metallic regimes ($E_g$ < 1.5 eV).

As a general trend, increasing the substitution concentration from 6.25% to 25% leads to a gradual narrowing of the bandgap across almost all alloys. However, this decrease does not always happen monotonically. A clear spatial dependence is observed across the periodic table: elements situated in the middle of the 3*d*, 4*d* and 5*d* rows (such as Cr, Mn, Fe, Co, Mo, Tc, Ru, Rh, Pd, Re, Os, Ir, Pt) exhibit extremely small bandgaps. In Figure 2, the cells where the bandgap drops below the critical threshold of 1.5 eV are highlighted in pink, signifying a transition into a narrow-gap semiconductor or even metallic state. These elements are therefore excluded from FRAM applications by leakage alone at 12.5% substitution and above, independently of their structural response, which is discussed above. All calculated values of remanent polarization, *c*/*a* ratio and band gap width are presented in supplementary data.

By combined study of all calculated parameters (i.e. remanent polarization, *c*/*a* ratio, and bandgap width) we conclude that Sc, Nb and Ta emerge as the best candidates for aluminium substitution in ferroelectric AlN for FRAM applications. They balance the structural softening required to lower the coercive field while maintaining strong dielectric properties. To understand the quantum-mechanical origins of this success and the non-monotonic nature of their evolution, a detailed analysis of their density of states (DOS) is required.

## ELECTRONIC ORIGINS: DENSITY OF STATES (DOS) STUDY

To clarify the mechanisms behind the non-monotonic lattice evolution and the bandgap trends, density of states (DOS) calculations were carried out for the three leading candidates Sc, Nb and Ta. The results, with all energy scales aligned to the valence band maximum (VBM), are shown in Figure 3.

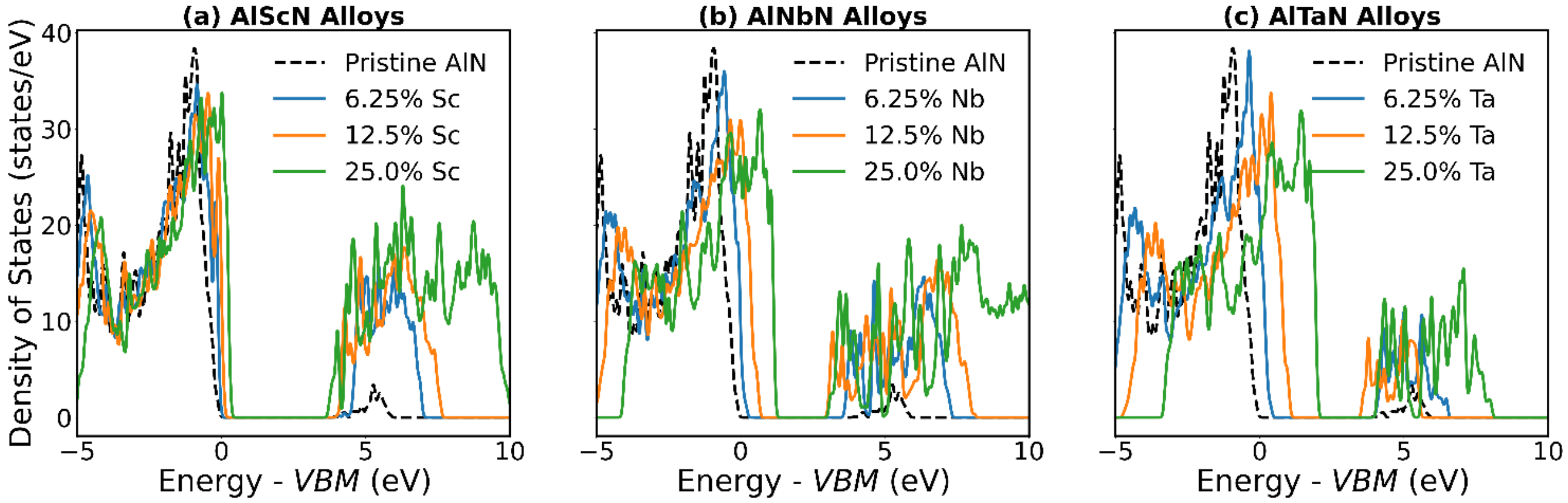

**Figure 3.** Evolution of the density of states (DOS) for (a) AlScN, (b) AlNbN, and (c) AlTaN alloys across different doping concentrations (6.25%, 12.5%, and 25%). Pristine AlN is plotted as a reference in each panel. All energy scales are aligned relative to the Valence Band Maximum (VBM = 0 eV).

In AlScN alloy (Fig. 3a), $Sc^{3+}$ substitutes $Al^{3+}$ isovalently and introduces no additional electrons. The bandgap remains free of mid-gap states at all three concentrations, and the wide-gap insulating character of the host is preserved ($E_g > 3.4$ eV), consistent with the established use of Sc as the reference dopant for ferroelectric AlN.

Nb and Ta substitute heterovalently: in their 5+ oxidation state each dopant contributes two excess electrons, which give rise to additional states inside the host bandgap (Fig. 3b, c). At 6.25% substitution these states appear as a sharp, narrow and isolated peak located above the VBM (centered at 3.60 eV and 4.09 eV for Nb and Ta, respectively), i.e. 0.66 eV and 0.17 eV below the conduction band edge of pristine AlN (4.26 eV) calculated at the same level of theory. The narrow width of the peak indicates that the dopants are spatially isolated and that the excess charge remains localized on individual dopant sites. The occupied states are *d*-derived and, under the $C_{3v}$ symmetry of the wurtzite cation position, predominantly in-plane in character; their occupation expands the lateral parameter *a* while leaving *c* comparatively unaffected, so that the *c*/*a* ratio decreases. This is the regime in which Nb and Ta are attractive for FRAM: lattice softening is achieved without introducing a conducting channel. The same argument accounts for the behaviour of the mid-row 3*d* dopants: once exchange splitting distributes the *d* electrons over both spin channels, the occupation of the low-lying in-plane states is no longer favoured, and the softening mechanism operative for Nb and Ta is lost. It should be noted that the role of *d* electrons in the present case differs from that established for binary wurtzite compounds with cations of fully occupied *d* shells, where a deeper *d* level weakens the $d(z^2)$-$p(z)$ hybridization along the polar axis and thereby raises the *c*/*a* ratio [11]. Here the dopant *d* states are only partially occupied and lie within the host bandgap, so it is their occupation rather than their depth below the valence band that governs the lattice response.

At 12.5% the impurity peak remains sharp and isolated, showing that the excess electrons are still localized, yet the lattice response reverses sharply and *c*/*a* expands by +3.21% (Nb) and +3.11% (Ta) relative to pristine AlN. Since the reported values correspond to the lowest-energy configuration among all symmetry-inequivalent dopant arrangements, this expansion is an intrinsic property of the composition rather than an artefact of a particular substitutional pattern.

At 25% the isolated peaks disappear and merge into a continuous, high-intensity band contiguous with the conduction band edge. This signals direct overlap of d orbitals on neighbouring dopants and delocalization of the excess electrons across the lattice, i.e. a transition to metallic-like electronic behaviour, and it coincides with the secondary compression of *c*/*a* to 1.616 (Nb) and 1.635 (Ta). A comparable non-monotonic evolution of the lattice parameters has been reported experimentally for Ta-substituted AlN thin films [31]. It should be emphasized that the 25% composition is of interest here only as evidence for the origin of the non-monotonic structural trend: the loss of the bandgap makes it unsuitable for memory operation, and the useful window for Nb and Ta lies at the low-substitution end, where localized impurity states coexist with a retained gap.

## CONCLUSIONS

Using *ab initio* calculations, we systematically evaluated the impact of 3*d*, 4*d* and 5*d* transition metal dopants on the structural, ferroelectric and electronic properties of wurtzite AlN at 6.25%, 12.5% and 25% substitution. The remanent polarization stays within 1.11-1.43 $C/m^2$ across all insulating configurations, so dopant selection is governed by the *c*/*a* ratio that determines the coercive field and the bandgap rather than by the polarization magnitude. Sc remains the reference dopant, while Nb and Ta are competitive and more cost-efficient alternatives at low substitution, where localized in-plane *d* states soften the lattice without closing the gap. The non-monotonic concentration dependence predicted for Nb, Ta, Zr and Y matches the lattice-parameter evolution reported experimentally for Ta-substituted AlN films. Spin-polarized calculations show that the mid-row 3*d* elements offer no usable window for coercive-field engineering, and that treating them without spin polarization overestimates their promise. More generally, lattice softening in AlN proceeds through the occupation of dopant *d* states, yet the same states narrow the bandgap and, once spin-split, cease to soften the lattice at all. This is why useful dopants remain confined to the early transition metals, where the *d* occupancy is low and the local moments vanish.

## FUNDING

Current work was financially supported by the Russian Science Foundation (Project No. 25-19-00493).

## CONFLICT OF INTEREST

The authors declare that they have no conflicts of interest.

## REFERENCES

1. Fichtner, S., Wolff, N., Lofink, F., Kienle, L., Wagner, B., 2019. AlScN: A III-V semiconductor based ferroelectric. Journal of Applied Physics 125. https://doi.org/10.1063/1.5084945
2. Yasuoka, S., Shimizu, T., Tateyama, A., Uehara, M., Yamada, H., Akiyama, M., Hiranaga, Y., Cho, Y., Funakubo, H., 2020. Effects of deposition conditions on the ferroelectric properties of $(Al_{1-x}Sc_x)N$ thin films. Journal of Applied Physics 128. https://doi.org/10.1063/5.0015281

3. Mikheev, V., Kondratyuk, E., Chouprik A 2022 Retention model and express retention test of ferroelectric $HfO_2$-based memory Phys. Rev. Appl. 18 064084 https://doi.org/10.1103/PhysRevApplied.18.064084
4. Shannigrahi, S.R., Jang, H.M., 2001. Fatigue-free lead zirconate titanate-based capacitors for nonvolatile memories. Applied Physics Letters 79, 1051–1053. https://doi.org/10.1063/1.1392970
5. Zhang, Y., Zhu, Q., Tian, B., Duan, C., 2024. New-generation ferroelectric AlScN materials. Nano-Micro Letters 16, 227. https://doi.org/10.1007/s40820-024-01441-1
6. Hasegawa, K., Shimizu, T., Ohsawa, T., Sakaguchi, I., Ohashi, N., 2023. Full polarization reversal at room temperature in unsubstituted AlN. Applied Physics Letters 123. https://doi.org/10.1063/5.0174236
7. Yang, F., 2024. Physics of Ferroelectric Wurtzite $Al_{1-x}Sc_xN$ Thin Films. Advanced Electronic Materials 11. https://doi.org/10.1002/aelm.202400279
8. Zhang, S., Fu, W.Y., Holec, D., Humphreys, C.J., Moram, M.A., 2013. Elastic constants and critical thicknesses of ScGaN and ScAlN. Journal of Applied Physics 114. https://doi.org/10.1063/1.4848036
9. Yazawa, K., Drury, D., Zakutayev, A., Brennecka, G.L., 2021. Reduced coercive field in epitaxial thin film of ferroelectric wurtzite $Al_{0.7}Sc_{0.3}N$. Applied Physics Letters 118. https://doi.org/10.1063/5.0043613
10. Yasuoka, S., Mizutani, R., Ota, R., Shiraishi, T., Shimizu, T., Uehara, M., Yamada, H., Akiyama, M., Funakubo, H., 2022. Tunable ferroelectric properties in wurtzite $(Al_{0.8}Sc_{0.2})N$ via crystal anisotropy. ACS Applied Electronic Materials 4, 5165–5170. https://doi.org/10.1021/acsaelm.2c00999
11. Zhou, Z., Xue, K.-H., Huang, J., Yu, H., Yang, S., Liu, S., Wang, Y., Miao, X., 2024. On the material genome of wurtzite ferroelectrics. arXiv preprint arXiv:2410.02471. https://doi.org/10.48550/arXiv.2410.02471
12. Yu, X., Zhang, C., Tang, R., Long, X., Wang, Y., Gong, T., Yang, Y., Jiang, P., Wei, W., Luo, Q., 2025. Influence of Sc, Y, and B dopants on structural distortions and polarization switching in AlN-based ferroelectrics. Journal of Applied Physics 138, 064102. https://doi.org/10.1063/5.0278485
13. Startt, J., Quazi, M., Sharma, P., Vazquez, I., Poudyal, A., Jackson, N., Dingreville, R., 2023. Unlocking AlN piezoelectric performance with earth-abundant dopants. Advanced Electronic Materials 9. https://doi.org/10.1002/aelm.202201187
14. Hayden, J., Hossain, M.D., Xiong, Y., Ferri, K., Zhu, W., Imperatore, M.V., Giebink, N., Trolier-McKinstry, S., Dabo, I., Maria, J.-P., 2021. Ferroelectricity in boron-substituted aluminum nitride thin films. Physical Review Materials 5. https://doi.org/10.1103/physrevmaterials.5.044412
15. Savant, C., Gund, V., Nomoto, K., Maeda, T., Jadhav, S., Lee, J., Ramesh, M., Kim, E., Nguyen, T.-S., Chen, Y.-H., Casamento, J., Rana, F., Lal, A., Xing, H.G., Jena, D., 2024. Ferroelectric AlBN films by molecular beam epitaxy. Applied Physics Letters 125. https://doi.org/10.1063/5.0181217
16. Zhang, J., Pan, Z., Guo, Y., Chu, M., Khokhar, M.Q., Yi, J., 2025. $Al_{1-x}B_xN$ Ferroelectric Thin Films: Research Progress and Prospects for Engineering Applications. Physica Status Solidi (A) 223. https://doi.org/10.1002/pssa.202500561

17. Zhao, Y., Zhou, Y., Ouyang, Y., Tao, X., 2026. Effect of doping on the ferroelectric properties of wurtzite $Al_{1-x}Y_xN$. Applied Physics Letters 128, 042901. https://doi.org/10.1063/5.0303096
18. Safaltin, S., Nayak, S.K., Alpay, S.P., 2026. Tailoring dielectric, piezoelectric and ferroelectric properties in aluminum scandium nitride through doping. Acta Materialia 308, 122040. https://doi.org/10.1016/j.actamat.2026.122040
19. Giannozzi, P., Baroni, S., Bonini, N., Calandra, M., Car, R., Cavazzoni, C., Ceresoli, D., Chiarotti, G.L., Cococcioni, M., Dabo, I., Corso, A.D., De Gironcoli, S., Fabris, S., Fratesi, G., Gebauer, R., Gerstmann, U., Gougoussis, C., Kokalj, A., Lazzeri, M., Martin-Samos, L., Marzari, N., Mauri, F., Mazzarello, R., Paolini, S., Pasquarello, A., Paulatto, L., Sbraccia, C., Scandolo, S., Sclauzero, G., Seitsonen, A.P., Smogunov, A., Umari, P., Wentzcovitch, R.M., 2009. QUANTUM ESPRESSO: a modular and open-source software project for quantum simulations of materials. Journal of Physics Condensed Matter 21, 395502. https://doi.org/10.1088/0953-8984/21/39/395502
20. Perdew, J.P., Burke, K., Ernzerhof, M., 1996. Generalized gradient approximation made simple. Physical Review Letters 77, 3865–3868. https://doi.org/10.1103/physrevlett.77.3865
21. Blöchl, P.E., 1994. Projector augmented-wave method. Physical Review. B, Condensed Matter 50, 17953–17979. https://doi.org/10.1103/physrevb.50.17953
22. Ong, S.P., Richards, W.D., Jain, A., Hautier, G., Kocher, M., Cholia, S., Gunter, D., Chevrier, V.L., Persson, K.A., Ceder, G., 2012. Python Materials Genomics (pymatgen): A robust, open-source python library for materials analysis. Computational Materials Science 68, 314–319. https://doi.org/10.1016/j.commatsci.2012.10.028
23. Von Barth, U., Hedin, L., 1972. A local exchange-correlation potential for the spin polarized case. i. Journal of Physics C Solid State Physics 5, 1629–1642. https://doi.org/10.1088/0022-3719/5/13/012
24. Qin, H., He, N., Han, C., Zhang, M., Wang, Y., Hu, R., Wu, J., Shao, W., Saadi, M., Zhang, H., Hu, Y., Liu, Y., Wang, X., Tong, Y., 2024. Perspectives of Ferroelectric Wurtzite AlScN: Material characteristics, preparation, and applications in advanced memory devices. Nanomaterials 14, 986. https://doi.org/10.3390/nano14110986
25. Clima, S.; Wouters, D. J.; Adelmann, C.; Schenk, T.; Schroeder, U.; Jurczak, M.; Pourtois G. Identification of the ferroelectric switching process and dopant-dependent switching properties in orthorhombic $HfO_2$: A first principles insight. Appl. Phys. Lett. 104 (9), 092906 (2014). https://doi.org/10.1063/1.4867975
26. Cao, T., Ren, G., Shao, D.-F., Tsymbal, E.Y., Mishra, R., 2023. Stabilizing polar phases in binary metal oxides by hole doping. Physical Review Materials 7. https://doi.org/10.1103/physrevmaterials.7.044412
27. Sawyerr, F., Sun, Y., Zhang, Z., Jia, K., Lv, S., Hu, Q., Shi, S., Huang, Q., Zhang, X., Fan, X., Liu, L., Wen, S., Wen, Z., Cao, T., Chen, J., 2026. Orientation-Confinement-Engineered Stabilization of Ferroelectricity in $HfO_2$ toward Maximum Polarization. Advanced Science 13, e21730. https://doi.org/10.1002/advs.202521730
28. Zhao, Q.X., Geng, B., Guan, L., Li, Q., Ma, J.K., Wei, D.Y., Guo, J.X., Liu, B.T., 2009. First-Principles Study of the Electronic Structure and Spontaneous Polarization of $PbZr_{0.4}Ti_{0.6}O_3$ Advanced Materials Research 79–82, 1249–1252. https://doi.org/10.4028/www.scientific.net/amr.79-82.1249

29. Oka, K., Kasai, H., 2014. Prediction of ferroelectric behavior for PZT and Pb-Substituted materials. e-Journal of Surface Science and Nanotechnology 12, 40–44. https://doi.org/10.1380/ejssnt.2014.40
30. Fu, D., Sogen, S., Suzuki, H., 2024. Intrinsic piezoelectricity of PZT. ACS Applied Electronic Materials 6, 6384–6390. https://doi.org/10.1021/acsaelm.4c00862
31. Liu, H., Zeng, F., Tang, G., Pan, F., 2013. Enhancement of piezoelectric response of diluted Ta doped AlN. Applied Surface Science 270, 225–230. https://doi.org/10.1016/j.apsusc.2013.01.005

**Table S1.** Remanent polarization, *c/a* ratio and electronic bandgap of transition-metal-substituted wurtzite AlN at 6.25%, 12.5% and 25% substitution. Values correspond to the lowest-energy configuration among all symmetry-inequivalent dopant arrangements. *c/a* change is given relative to pristine AlN. Blank entries denote configurations that did not converge; "n/a" denotes cases where the Berry-phase polarization could not be evaluated.

## 6.25% substitution

| atom | Sc | Ti | V | Cr | Mn | Fe | Co | Ni | Cu | Zn | pure AlN |
|---|---|---|---|---|---|---|---|---|---|---|---|
| polarization, C/m2 | 1,293995 | 1,2881662 | | 1,305225 | 1,3200253 | 1,3218663 | 1,325825 | | 1,329435 | 1,32244 | 1,3247202 |
| c/a | 1,557591 | 1,59747 | n/a | 1,601314 | 1,60784505 | 1,603325 | 1,60348627 | n/a | 1,613282 | 1,613593 | 1,60332013 |
| c/a change, % | -2,852152 | -0,3648758 | | -0,1251233 | 0,28222169 | 0,0003039 | 0,01036265 | | 0,62132773 | 0,64072498 | 0 |
| bandgap, eV | 4,1112 | 3,5747 | | 1,4515 | 0 | 1,197 | 1,6762 | | 3,291 | 3,633 | 4,26 |
| | **Y** | **Zr** | **Nb** | **Mo** | **Tc** | **Ru** | **Rh** | **Pd** | **Ag** | **Cd** | |
| polarization, C/m2 | 1,285318 | 1,243471 | 1,2538127 | 1,277461 | 1,2537 | 1,3176754 | 1,297448 | 1,3205 | 1,31168 | 1,30317 | |
| c/a | 1,601602 | 1,595019 | 1,590086 | 1,594543 | 1,58962 | 1,612431 | 1,596718 | 1,616961 | 1,623862 | 1,619035 | |
| c/a change, % | -0,1071606 | -0,5177461 | -0,8254202 | -0,5474345 | -0,8544848 | 0,56825037 | -0,4117785 | 0,85078908 | 1,28120842 | 0,98014565 | |
| bandgap, eV | 3,5713 | 3,9131 | 3,4218 | 2,2358 | 2,3726 | 0,6169 | 1,0044 | 2,1914 | 2,9077 | 3,0094 | |
| | **La** | **Hf** | **Ta** | **W** | **Re** | **Os** | **Ir** | **Pt** | **Au** | **Hg** | |
| polarization, C/m2 | 1,3119 | 1,272475 | 1,251843 | | 1,2589 | 1,327913 | 1,284538 | 1,326174 | 1,323703 | 1,3135621 | |
| c/a | 1,62387 | 1,602571 | 1,594027 | n/a | 1,586278 | 1,618906 | 1,592408 | 1,620643 | 1,633793 | 1,625051 | |
| c/a change, % | 1,28170739 | -0,0467235 | -0,5796177 | | -1,0629273 | 0,97209985 | -0,6805957 | 1,08043754 | 1,90061062 | 1,35536704 | |
| bandgap, eV | 2,9077 | 3,9234 | 3,661 | | 2,4563 | 0,6598 | 1,0976 | 1,8245 | 2,5863 | 2,4521 | |

## 12.5% substitution

| atom | Sc | Ti | V | Cr | Mn | Fe | Co | Ni | Cu | Zn |
|---|---|---|---|---|---|---|---|---|---|---|
| polarization, C/m2 | 1,2701 | 1,2992392 | | 1,3888007 | 1,3111872 | 1,3351291 | 1,3290786 | | | 1,2911006 |
| c/a | 1,598554 | 1,6074436 | | 1,61343148 | 1,6189089 | 1,60460514 | 1,6023027 | n/a | n/a | 1,631092 |
| c/a change, % | -0,2972661 | 0,25718336 | | 0,63065094 | 0,97228095 | 0,08014669 | -0,0634578 | | | 1,73214769 |
| bandgap, eV | 3,9517 | 3,6354 | n/a | 1,2508 | 0 | 0,0606 | 0,5044 | | | 3,501 |
| | **Y** | **Zr** | **Nb** | **Mo** | **Tc** | **Ru** | **Rh** | **Pd** | **Ag** | **Cd** |
| polarization, C/m2 | 1,202792 | 1,3176761 | 1,296 | 1,312674 | 1,228964 | 1,224536 | | 1,316164 | 1,32101 | 1,2722961 |
| c/a | 1,60859 | 1,636465 | 1,654861 | 1,632291 | 1,615789 | 1,594301 | n/a | 1,648417 | 1,652804 | 1,648938 |
| c/a change, % | 0,32868499 | 2,06726479 | 3,21463391 | 1,80693001 | 0,77769076 | -0,5625282 | | 2,81271792 | 3,08633764 | 2,84521299 |
| bandgap, eV | 3,0559 | 2,6145 | 2,4019 | 1,5181 | 2,1137 | 1,3538 | | 1,2658 | 2,2138 | 2,7364 |
| | **La** | **Hf** | **Ta** | **W** | **Re** | **Os** | **Ir** | **Pt** | **Au** | **Hg** |
| polarization, C/m2 | 1,32101 | 1,2639291 | 1,2427 | | | 1,2976139 | 1,3467 | 1,31716 | 1,309779 | 1,2793302 |
| c/a | 1,652801 | 1,607788 | 1,65313 | n/a | n/a | 1,647702 | 1,669032 | 1,653031 | 1,66207 | 1,655567 |
| c/a change, % | 3,08615053 | 0,27866378 | 3,10667045 | | | 2,76812296 | 4,09848735 | 3,10049576 | 3,6642634 | 3,25866754 |
| bandgap, eV | 2,2138 | 3,0739 | 2,5078 | | | 1,3186 | 0,9561 | 0,5076 | 1,5117 | 1,9956 |

## 25% substitution

| atom | **Sc** | **Ti** | **V** | **Cr** | **Mn** | **Fe** | **Co** | **Ni** | **Cu** | **Zn** |
|---|---|---|---|---|---|---|---|---|---|---|
| polarization, C/m2 | 1,209303 | 1,2318306 | | | 1,3015166 | 1,3072147 | 1,3206573 | | 1,338255 | 1,323013 |
| c/a | 1,586357 | 1,616283 | n/a | n/a | 1,61911676 | 1,60174748 | 1,5975272 | n/a | 1,627455 | 1,613642 |
| c/a change, % | -1,058 | 0,80850183 | | | 0,98524519 | -0,0980867 | -0,3613085 | | 1,5053059 | 0,64378114 |
| bandgap, eV | 3,4176 | 2,3066 | | | 0 | 0 | 0 | | 2,4006 | 3,0776 |
| | **Y** | **Zr** | **Nb** | **Mo** | **Tc** | **Ru** | **Rh** | **Pd** | **Ag** | **Cd** |
| polarization, C/m2 | 1,159461 | 1,1128463 | 1,182667 | 1,272003 | 1,29734 | 1,4221398 | 1,4047 | 1,3499768 | 1,294091 | 1,138522 |
| c/a | 1,594562 | 1,598678 | 1,616216 | 1,651028 | 1,68259 | 1,686044 | 1,673876 | 1,664114 | 1,647049 | 1,647274 |
| c/a change, % | -0,5462495 | -0,2895322 | 0,804323 | 2,97556749 | 4,94410762 | 5,15953559 | 4,40061042 | 3,79174885 | 2,72739497 | 2,74142835 |
| bandgap, eV | 2,5259 | 1,9436 | 1,9207 | 1,0052 | 0,3217 | 0 | 0 | 0,1166 | 1,7202 | 2,0211 |
| | **La** | **Hf** | **Ta** | **W** | **Re** | **Os** | **Ir** | **Pt** | **Au** | **Hg** |
| polarization, C/m2 | 1,29408 | 1,1516554 | 1,223713 | | 1,412 | 1,427406 | 1,4226 | 1,3736967 | 1,2577 | 1,148922 |
| c/a | 1,64706 | 1,608472 | 1,635254 | n/a | 1,704022 | 1,697852 | 1,694565 | 1,673167 | 1,639699 | 1,656289 |
| c/a change, % | 2,72808105 | 0,32132526 | 1,99173403 | | 6,2808338 | 5,89600735 | 5,69099527 | 4,35638968 | 2,26897124 | 3,30369909 |
| bandgap, eV | 1,72 | 2,1143 | 2,0631 | | 0,1876 | 0,1148 | 0,6186 | 0,3705 | 0,9273 | 1,1843 |